\documentclass[runningheads]{svmult}

\usepackage{makeidx}   % allows index generation
\usepackage{graphicx}  % standard LaTeX graphics tool
\usepackage{subeqnar}  % subnumbers individual equations
\usepackage{multicol}  % used for the two-column index
\usepackage{physprbb}  % modified textarea for proceedings,
\makeindex             % used for the subject index
\begin{document}
\title*{The Influence of Black Hole Mass and Accretion Rate 
on the FRI/FRII Radio Galaxy Dichotomy}
\toctitle{The Influence of Black Hole Mass and Accretion Rate 
on the FRI/FRII Radio Galaxy Dichotomy} 
%\protect\newline in the Particle Deflection Plane}
% allows explicit linebreak for the table of content
%
%
\titlerunning{Black Hole Mass and Accretion Rate of Radio 
Galaxies}
% allows abbreviation of title, if the full title is too long
% to fit in the running head
%
\author{Margrethe Wold\inst{1,2}
\and Mark Lacy\inst{2}
\and Lee Armus\inst{2}
}
\authorrunning{Wold et al.}
\institute{European Southern Observatory, D-85748 Garching 
           bei M{\"u}nchen, Germany
\and Spitzer Science Center/Caltech, 1200 E.\ California Blvd.,
     Pasadena, CA 91125}

\maketitle              % typesets the title of the contribution

\begin{abstract}
%70 and at most 150 words
We use medium resolution optical spectra of 3CR radio 
galaxies to estimate their black hole masses and accretion 
rates. Black hole masses are found from central stellar velocity
dispersions, and accretion rates are derived from narrow 
emission-line luminosities. The sample covers both 
Fanaroff-Riley (FR) classes; the more powerful FRIIs and the 
less powerful FRIs. We find that FRIs and FRIIs separate in 
diagrams of radio luminosity and narrow-line luminosity 
versus black hole mass. This suggests that, at a given 
black hole mass, the FRIIs accrete more efficiently, 
or accrete more matter, than FRIs.
\end{abstract}

\section{Introduction}

The classification of radio galaxies into FRIs and FRIIs is morphological 
in origin. Whereas FRIIs have knotty jets and radio lobes 
that end in bright hot-spots, the less powerful 
FRIs have smooth jets and large lobes 
that grow dimmer outward. The FRI/FRII transition is relatively 
sharp in terms of radio luminosity, with sources at 
$\log P_{\rm 178 MHz}>25$ (W\,Hz$^{-1}$\,sr$^{-1}$) being almost 
exclusively FRIIs, and sources at lower radio luminosities possessing 
FRI morphology \cite{fr74}. Interestingly, the FRI/FRII transition 
radio luminosity, $L_{\rm rad}$, is an increasing function of the 
optical luminosity, $L_{\rm opt}$, of the host galaxy,
$L_{\rm rad} \propto L_{\rm opt}^{2}$ \cite{lo96}.
The separation of FRIs and FRIIs in the radio-optical luminosity 
plane is surprisingly sharp, and several explanations have been offered
including environmental differences \cite{deyoung93,gopal01}, 
different physical processes \cite{meier99} and differences in AGN 
parameters such as black hole mass and accretion rate 
\cite{baum95,gc01}. 

\section{Results}

Here we investigate the influence of black hole mass and accretion
rate on the FRI/FRII dichotomy. 
Spectra were obtained 
with the Double Spectrograph on the Palomar 5m, and
sources having $P_{\rm 178 MHz} \leq 10^{26.5}$
W\,Hz$^{-1}$\,sr$^{-1}$ (assuming $H_{0}=70$ km\,s$^{-1}$\,Mpc$^{-1}$, 
$\Omega_{m}=0.2$, $\Omega_{\Lambda}=0.8$) were selected. 

The spectra cover the regions around the Ca H\&K $\lambda\lambda$3934,3969  
{\AA} and the MgI$b$ $\lambda$5175 {\AA} absorption. Template stars 
were observed and broadened with Gaussians and fit to 
the galaxy spectra using a direct fitting routine \cite{barth02}. 
The best-fit velocity dispersions were converted to black hole masses
using the $M_{\rm bh}$--$\sigma_{*}$ relation \cite{fm00,gebhardt00}.  
From the literature, but also from wide-slit spectra, we obtain 
the luminosity in the narrow [OII]3727 and/or [OIII]5007 {\AA} emission lines.

\begin{figure}[b]
\begin{center}
\includegraphics[width=0.93\textwidth]{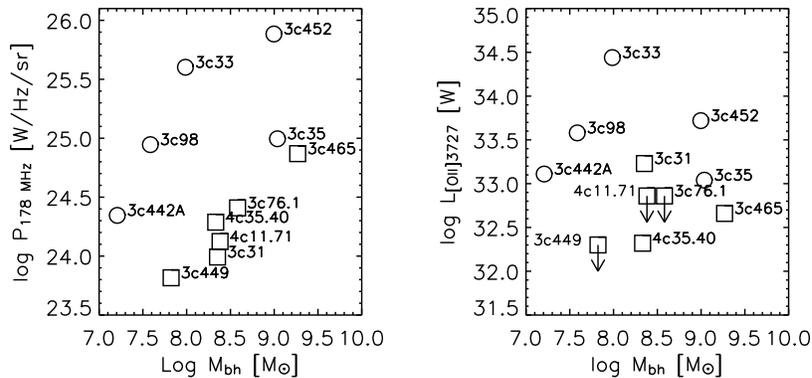}
\end{center}
\caption[]{Radio luminosity (left) and narrow-line [OII]3727 luminosity
(right) as a function of black hole mass. FRI sources are plotted with 
squares, and FRIIs with circles.}
\label{fig1}
\end{figure}

In Fig.~\ref{fig1}, we show the radio and the narrow-line [OII] 
luminosity as a function of black hole mass for some of the galaxies
in our sample. FRIs and FRIIs separate 
fairly well, populating different regions of the diagrams. The [OII]
narrow-line luminosity probes the total luminosity in the 
narrow-line region \cite{willott99}, and taking into account a 
covering factor, the narrow line luminosity scales with 
the photoionizing luminosity, and hence the accretion rate.
The separation of FRIs and FRIIs
in these diagrams may therefore imply that, at a given black hole mass,
FRIs accrete less efficiently (or accrete less matter) than FRIIs. 
Sources that do not follow the general separation so well may be
low-excitation radio galaxies \cite{laing94}. One such example is
3C 35, which we find to have much weaker [OIII] than [OII].
In this case the environment, rather than accretion rate and black hole mass, 
may have determined the radio morphology of the source. 

%INDEX%%%%%%%%%%%%%%%%%%%%%%%%%%%%%%%%%%%%%%%%%%%%%%%%%%%%%%%%%%%%%%%
% Please check with the editor of your book whether he plans to
% include a "mutual" subject index - if so, please code your entries
% in the standard syntax. For your own purposes you may print your
% "personal" index by using the following commands:
%
%\clearpage
%\addcontentsline{toc}{section}{Index}
%\flushbottom
%\printindex
%%%%%%%%%%%%%%%%%%%%%%%%%%%%%%%%%%%%%%%%%%%%%%%%%%%%%%%%%%%%%%%%%%%%%

\end{document}